\documentclass[reprint,showpacs,prepintnumbers,amsmath,amssymb,aps,twocolumn]{aastex61}
\usepackage{amsmath,amssymb}
\usepackage{graphicx}
\usepackage{color}
\usepackage{comment}
\usepackage{bbold}
\usepackage{soul,color}
\sethlcolor{white}
\usepackage{booktabs}
\usepackage{array}
\usepackage{hyperref}
\hypersetup{
    colorlinks=true,       
    linkcolor=blue,          
    citecolor=blue,        
    filecolor=blue,      
    urlcolor=blue           
}

\def\Al{$^{26}$Al}
\def\Cl{$^{34}$Cl}
\def\Kr{$^{85}$Kr}
\def\Mg{$^{26}$Mg}

\begin{document}

\title{Using Steady State Behavior to Assess Treatments of Nuclear Isomers in Astrophysical Environments}

\author{G. Wendell Misch}
\email{Preferred: wendell.misch@gmail.com}
\email{wendell@sjtu.edu.cn}
\affiliation{Department of Physics, School of Physics and Astronomy, Shanghai Jiao Tong University, Shanghai 200240, China}
\affiliation{Collaborative Innovation Center of IFSA, Shanghai Jiao Tong University, Shanghai 200240, China}


\author{Surja K. Ghorui}
\email{surja@sjtu.edu.cn}
\affiliation{Department of Physics, School of Physics and Astronomy, Shanghai Jiao Tong University, Shanghai 200240, China}

\author{Yang Sun}
\email{sunyang@sjtu.edu.cn}
\affiliation{Department of Physics, School of Physics and Astronomy, Shanghai Jiao Tong University, Shanghai 200240, China}
\affiliation{Collaborative Innovation Center of IFSA, Shanghai Jiao Tong University, Shanghai 200240, China}
\affiliation{Institute of Modern Physics, Chinese Academy of Sciences, Lanzhou 730000, China}
\affiliation{China Institute of Atomic Energy, P.O. Box 275(10), Beijing 102413, China}

\begin{abstract}
Differing reaction rates of long-lived nuclear states can force the level occupations out of thermal equilibrium, causing calculations of overall rates which rely on thermal equilibrium to be inaccurate.  Therefore, nucleosynthesis calculations which include nuclei with isomers must use techniques that do not assume thermal equilibrium, and it is imperative that such techniques appropriately account for transitions between the ground and isomeric states via higher-lying levels.  We develop a formalism to compute the steady state occupations of nuclear levels and apply it to the examples {\Al}, {\Cl}, and {\Kr}.  We show that this approach is useful both for assessing the required number of nuclear levels and for determining the temperature above which thermal equilibrium rates are appropriate.
\end{abstract}


\section{\label{intro}Introduction}

In astrophysical nuclear reaction calculations, nuclear isomers present a particular challenge.  An isomer is an excited nuclear state with a lifetime much longer than typical excited states.  Known isomers in nuclei span the range of lifetimes from $10^{15}$ years in $^{180}$Ta -- much longer than the accepted age of the universe -- to an informal rule of thumb on the lower side of approximately 1 ns.  Isomers arise from nuclear structure effects (spin-trap, shape change, K quantum number, etc.) that inhibit $\gamma$-decay to lower energy levels \citep{wd:1999}.

In most isotopes, thermally-driven electromagnetic transitions are fast and keep the nuclear state occupations in thermal equilibrium.  Indeed, thermal $\beta$-decay rates and neutrino spectra are typically computed under the assumption of a Boltzmann distribution of level occupations \citep{ffn:1982b,oda-etal:1994,lm:2001,msf:2018}, with the total $\beta$-decay rate of the nucleus given by a thermal weighting of the decay rates of the individual states.

\begin{equation}
    \lambda^{\beta} = \sum\limits_i \lambda_{i}^{\beta}n_i
    \label{eq:beta_inst}
\end{equation}
The sum is over all nuclear states $i$, $\lambda_{i}^{\beta}$ is the $\beta$-decay rate of state $i$, and $n_i$ is the occupation fraction of state $i$.

\begin{equation}
    n_i = \frac{g_i}{G(T)}e^{-E_i/T}
    \label{eq:ni_therm}
\end{equation}
Here $g_i=2J_i+1$ is the degeneracy of state $i$, $J_i$ is the state's spin, $G(T)$ is the nuclear partition function at temperature $T$, and $E_i$ is the energy of nuclear state $i$.
 
The suppressed $\gamma$ transitions between an isomer and lower-lying levels, however, can cause these long-lived states to fall out of thermal equilibrium, particularly if the isomer's destruction rate is vastly different from the ground state (GS).  Essentially, if one destruction rate proceeds faster than the long-lived states can equilibrate, the rapidly-detroyed state will be depopulated relative to its thermal equilibrium occupation.  This in turn results in a deviation of the total destruction rate from its thermal equilibrium value.  Hence, when an isotope has a long-lived isomer sufficiently low in energy that it will have an appreciable thermal equilibrium occupation, it must be handled carefully in nucleosynthesis calculations.
 
One of the best-known examples of this situation is the $\beta$-decay of {\Al}, a radioisotope used as a cosmochronometer for early solar system studies and $\gamma$-ray astronomy.  The GS of {\Al} has a half-life against $\beta$-decay of $0.717$ Myr, but it also has a long-lived isomer at 228 keV.  This isomer has a super-allowed $\beta$-decay to the GS of {\Mg}, giving it a half-life of 6.35 s.  Thermal processes at low temperature populate this isomer slowly compared to its $\beta$-decay rate, so its occupation in medium will be significantly lower than the thermal Boltzmann value.  This results in a lower $\lambda^\beta$ than that computed from a Boltzmann distribution.  Because of its observational importance, the $\beta$-decay rates of {\Al} in a thermal bath have been studied extensively, including numerous approaches to computing an effective $\lambda^\beta$ \citep{wf:1980,coc2000,runkle2001,gupta2001,iliadis2011,reifarth}.

The techniques from previous studies consider differing numbers of nuclear levels in their calculations.  It is generally agreed that the number of included states can impact the results, as higher-lying levels act as intermediate states that facilitate ``communication'' between the GS and the isomer.  At sufficiently high temperature, however, thermal processes are fast enough to keep the nuclear levels in thermal equilibrium.  Above this equilibration temperature, effective treatments should in principle converge, and thermal destruction rates are appropriate.

This paper provides a method for determining an appropriate number of nuclear levels to include in nucleosynthesis calculations.  This technique also provides the equilibration temperature.  After some brief general comments on isomers in section \ref{sec:isomers}, we show how to express the long term ($t\rightarrow\infty$) steady state nuclear level occupations from which to compute total destruction rates.  This formulation, detailed in section \ref{sec:evolution}, includes the effects of producing nuclei in any configuration of initial states and uses generalized destruction rates.  We give example results for {\Al}, {\Kr}, and {\Cl} in section \ref{sec:results}, and we discuss the results with some concluding remarks in section \ref{sec:concl}.

\section{Comments on Nuclear Isomers \label{sec:isomers}}

Often discussed in the literature are three mechanisms leading to nuclear isomerism \citep{wd:1999}, although new types of isomer may be possible in exotic nuclei \citep{hasegawa-etal:2005}.  It is difficult for an isomeric state to change its shape to match the states to which it is decaying, or to change its spin, or to change its spin orientation relative to an axis of symmetry.  These correspond to shape isomers, spin traps, and K-isomers, respectively.  In all of these cases, decay to the GS is strongly hindered either by an energy barrier or by the selection rules of transition, and isomer lifetimes can be remarkably long.

To give some examples of isomers, an $I^\pi = 0^+$ excited state in $^{72}$Kr has been found as a shape isomer \citep{bouchez-etal:2003}, a $12^+$ state in $^{98}$Cd has been understood as a spin trap \citep{blazhev-etal:2004}, and in $^{178}$Hf, there is a famous $16^+$, 31-year K-isomer \citep{sun-etal:2004} which has been a discussed for use as energy storage \citep{wc:2005}.  A laboratory analog of thermal processes facilitating isomer-to-ground transitions -- triggering the 75 keV $J^\pi=9^-$ $^{180m}$Ta isomer to de-excite by photon activation -- has been demonstrated \citep{belic-etal:1999}, and the detailed gamma-transition paths are well established \citep{wdc:2001}.

Detailed nuclear structure studies are at the heart of understanding the formation of nuclear isomers and their applications to various aspects of nuclear astrophysics \citep{hasegawa-etal:2011}.  Isomers may play a significant role in determining the abundances of the elements in the universe via their impact on various nucleosynthesis processes \citep{as:2005}, where an isomer of sufficiently long lifetime (probably longer than microseconds) can change the reaction paths and lead to a different set of elemental abundances.  \cite{sun-etal:2005} performed a rough comparison between two extreme possibilities in the rp-process (rapid proton capture process) for a reaction sequence calculated in the framework of a multi-mass-zone X-ray burst model.  In one case, the nuclear reactions were considered to proceed entirely through the ground state, while in the other case, all reactions proceeded through the isomer.  This study found pronounced differences.  The present work focuses on the illustrative example of $\beta$-decay, but the ideas are broadly applicable to other nuclear reaction channels.

\section{Evolution Equations \label{sec:evolution}}

The nuclear state abundances $N_i$ (the number of nuclei of a given species in state $i$) evolve according to the coupled differential equations \citep{wf:1980}
\begin{equation}
    \dot{N_i} = \sum\limits_{j}\left(\lambda_{ji}N_j - \lambda_{ij}N_i\right) - \lambda_{i}^{D}N_i + P_{i}
    \label{eq:Ni_dot}
\end{equation}
where $\lambda_{ij}$ is the internal transition (IT) rate from state $i$ to state $j$, $\lambda_{i}^{D}$ is the external destruction rate of state $i$ (e.g. $\beta$-decay), and $P_{i}$ is the production of state $i$ through all external channels.  The production rate has units of nuclei per unit time, as distinguished from the destruction rate units of nuclei per unit time per nucleus; it is expressed this way because production channels are generally independent of the abundance of the nucleus being produced.

We compute the $\lambda_{ij}$ using the spontaneous $\gamma$-decay rates $\lambda_{hl}^s$ from a higher state $h$ to a lower state $l$ \citep{coc2000}.

\begin{eqnarray}
    \lambda_{hl} = \lambda_{hl}^s (1+u) \\
    \lambda_{lh} = \frac{2J_h+1}{2J_l+1}\lambda_{hl}^s u \\
    u = \frac{1}{e^{(E_h-E_l)/T}-1}
\end{eqnarray}
The factor $u$ represents the effects of the thermal photon bath that stimulates transitions from $h$ to $l$ and induces transitions from $l$ to $h$.  As \cite{wf:1980} point out, there can be other thermal interactions which affect the $\lambda_{ij}$, but we consider here only the photon bath.

Writing equation \ref{eq:Ni_dot} in matrix form, we have
\begin{align}
    \mathbf{\dot{N}} = \left(\mathbf{\Lambda}^\mathbf{IT_{in}} - \mathbf{\Lambda^{IT_{out}}} -\mathbf{\Lambda^\mathbf{D}} \right)\mathbf{N} + \mathbf{P}
    \label{eq:Nvec_dot}
\end{align}
where $\mathbf{N}$ is the vector with components $N_i$, $\mathbf{\Lambda^{IT_{in/out}}}$ are the matrices of IT rates which feed into/out of nuclear states, $\mathbf{\Lambda}^\mathbf{D}$ is the diagonal matrix with elements $\lambda_{i}^{D}$, and $\mathbf{P}$ is the vector with components $P_{i}$.  The $\mathbf{\Lambda^{IT}}$ are constructed from the $\lambda_{ij}$; they are derived from the expression for the components of $\mathbf{\dot{N}}$ (equation \ref{eq:Ni_dot}).  If we consider $S$ distinct nuclear levels, they are given by
\begin{align}
    \mathbf{\Lambda^{IT_{in}}} &=
    \begin{bmatrix}
        0 & \lambda_{21} & \dots & \lambda_{S1} \\
        \lambda_{12} & 0 & \dots & \lambda_{S2} \\
        \vdots & \vdots & \ddots & \vdots \\
        \lambda_{1S} & \lambda_{2S} & \dots & 0
    \end{bmatrix} \\
    \mathbf{\Lambda^{IT_{out}}} &=
    \begin{bmatrix}
        \sum\limits_{i}\lambda_{1i} & 0 & \dots & 0 \\
        0 & \sum\limits_{i}\lambda_{2i} & \dots & 0 \\
        \vdots & \vdots & \ddots & \vdots \\
        0 & 0 & \dots & \sum\limits_{i}\lambda_{Si}
    \end{bmatrix}.
\end{align}

While equation \ref{eq:Nvec_dot} is completely general, it is not convenient for understanding the intrinsic behavior of the nucleus.  For this, it is helpful to use the nuclear level occupation fractions $\mathbf{n}$.
\begin{align}
    \mathbf{n}=\frac{\mathbf{N}}{N} \nonumber \\
    N \equiv \sum\limits_i N_i& ~\Rightarrow ~\sum\limits_i n_i = 1, ~~0\leq n_i\leq 1. \label{eq:n_definition}
\end{align}

From equations \ref{eq:Nvec_dot} and \ref{eq:n_definition}, we can derive an expression for $\mathbf{\dot{n}}$.
\begin{align}
    \dot{\mathbf{N}} &= \frac{d}{dt}N\mathbf{n} = \dot{N}\mathbf{n} + N\dot{\mathbf{n}} \nonumber \\
    \Rightarrow \mathbf{\dot{n}} &= \frac{1}{N}\mathbf{\dot{N}} - \frac{\dot{N}}{N}\mathbf{n} \nonumber \\
    &= \left(\mathbf{\Lambda}^\mathbf{IT_{in}} - \mathbf{\Lambda^{IT_{out}}} -\mathbf{\Lambda^\mathbf{D}} - \frac{\dot{N}}{N}\mathbb{1} \right)\mathbf{n} + \frac{\mathbf{P}}{N}
\end{align}
where $\mathbb{1}$ is the identity matrix of dimension $S$; we use the otherwise unnecessary identity matrix here so that the terms in parentheses have the same dimensions.  Using the fact that $N$ can only change via non-conservative interactions (production/destruction), we have
\begin{align}
    \dot{N} = \sum\limits_i\left( P_{i} - \lambda_{i}^{D}N_i\right) &= P - \lambda^{D} N \label{eq:N_dot} \\
    \lambda^{D} &\equiv \sum\limits_{i}\lambda_{i}^{D}n_i \nonumber
\end{align}
where $P=\sum\limits_i P_i$ is the total production of the nuclear species and $\lambda^{D}$ is the total destruction rate (cf. equation \ref{eq:beta_inst}).  This at last yields
\begin{widetext}
    \begin{equation}
        \mathbf{\dot{n}} = \left( \mathbf{\Lambda}^\mathbf{IT_{in}} - \mathbf{\Lambda^{IT_{out}}} - \mathbf{\Lambda^\mathbf{D}} + \left( \lambda^{D} - \frac{P}{N} \right) \mathbb{1} \right) \mathbf{n} + \frac{\mathbf{P}}{N}.
        \label{eq:nvec_dot}
    \end{equation}
    In component notation, this is
    \begin{equation}
        \dot{n}_i = \sum\limits_j \left( \lambda_{ji}n_j - \lambda_{ij}n_i \right) - \lambda_{i}^{D}n_i + \left( \lambda^{D} - \frac{P}{N} \right)n_i + \frac{P_i}{N}.
    \end{equation}
\end{widetext}

\subsection{Steady State Solutions \label{sec:steady_state}}

Absent changes in the environement, the nuclear level occupations will tend toward a steady state configuration which dictates, for example, the isotope's thermal $\beta$-decay rate.  The steady state condition is
\begin{align}
    \dot{\mathbf{n}} &= 0.
    \label{eq:ss_cond_n}
\end{align}
That is, the steady state occupation fraction of each nuclear state is time-independent and depends only on the IT and creation/destruction rates (which may themselves be functions of temperature, abundance of external reactants, etc.).

In the presence of production, we also have the long term requirement $\dot{N}=0$, that is, the total production and destruction rates reach equilibrium.  From equation \ref{eq:N_dot} we then have
\begin{equation}
    P = \lambda^D N.
    \label{eq:P_equals_D}
\end{equation}
Using a production fraction vector
\begin{equation}
    \mathbf{p} = \frac{\mathbf{P}}{P}
\end{equation}
and the steady state conditions \ref{eq:ss_cond_n} and \ref{eq:P_equals_D}, equation \ref{eq:nvec_dot} becomes
\begin{equation}
    \left( \mathbf{\Lambda}^\mathbf{IT_{in}} - \mathbf{\Lambda^{IT_{out}}} - \mathbf{\Lambda^\mathbf{D}} \right) \mathbf{n} + \lambda^D\mathbf{p} = 0,
    \label{eq:ss_prod}
\end{equation}
which is entirely independent of both $N$ and the total production $P$ and depends only on the production fractions $\mathbf{p}$ and the IT and destruction rates of the nucleus.  In the absence of production, equation \ref{eq:nvec_dot} in steady state becomes
\begin{equation}
    \left( \mathbf{\Lambda}^\mathbf{IT_{in}} - \mathbf{\Lambda^{IT_{out}}} - \mathbf{\Lambda^\mathbf{D}} + \lambda^D\mathbb{1} \right) \mathbf{n} = 0.
    \label{eq:ss_no_prod}
\end{equation}

Equations \ref{eq:ss_prod} and \ref{eq:ss_no_prod} coupled with the restrictions on $\mathbf{n}$ from equation \ref{eq:n_definition} constitute systems of equations which can be solved for the steady state values of $\mathbf{n}$ and $\lambda_{D}$.  In the presence of production, this will also allow for the computation of the steady state abundance of the nuclear species from equation \ref{eq:N_dot}.

\section{Results \label{sec:results}}

We applied the steady state techniques described above to three nuclei of astrophysical interest: {\Al}, {\Cl}, and {\Kr}.  Each of these isotopes has a long lived low-lying isomer that can affect the evolution of its abundance.  Furthermore, these isotopes are of interest in distinct astrophysical environments with different active nucleosynthesis processes.

\subsection{\Al}

{\Al} is an indicator of star formation, observed by $\gamma$ radiation from the daugher {\Mg} nuclei after it undergoes $\beta$ decay \citep{diehl1995}.  Its $\sim 1$ Myr vacuum lifetime against $\beta$ decay ensures that it decays slowly enough to be present well after a star's death, but not so slowly that it will diffuse broadly into the galactic medium and become uncorrelated with its production site.

For clarity, we index the nuclear levels by energy.  {\Al} has a long-lived isomer at 228 keV.  Its $\beta$-decay rate $\lambda_{228}^\beta = 0.11$ s$^{-1}$ is many orders of magnitude faster than the $\lambda_0^\beta = 3.1\times 10^{-14}$ s$^{-1}$ rate of the GS.  This implies two things.  First, when the occupation fraction of the isomer is $n_{228}\gtrsim 3\times 10^{-13}$, it dominates the GS's contribution to the total $\beta$-decay rate $\lambda^\beta$.  Second, because thermal transitions between ground and the isomer are extremely slow at low temperatures, the isomer tends to become depopulated via $\beta$ decay relative to its TE value, effectively decreasing $\lambda^\beta$.

Except where otherwise indicated, our calculations for {\Al} include the lowest 11 energy levels.  The IT and $\beta$-decay rates are identical to those in \cite{banerjee-etal:2018}.

\begin{figure}
    \centering
    \includegraphics[width=\columnwidth]{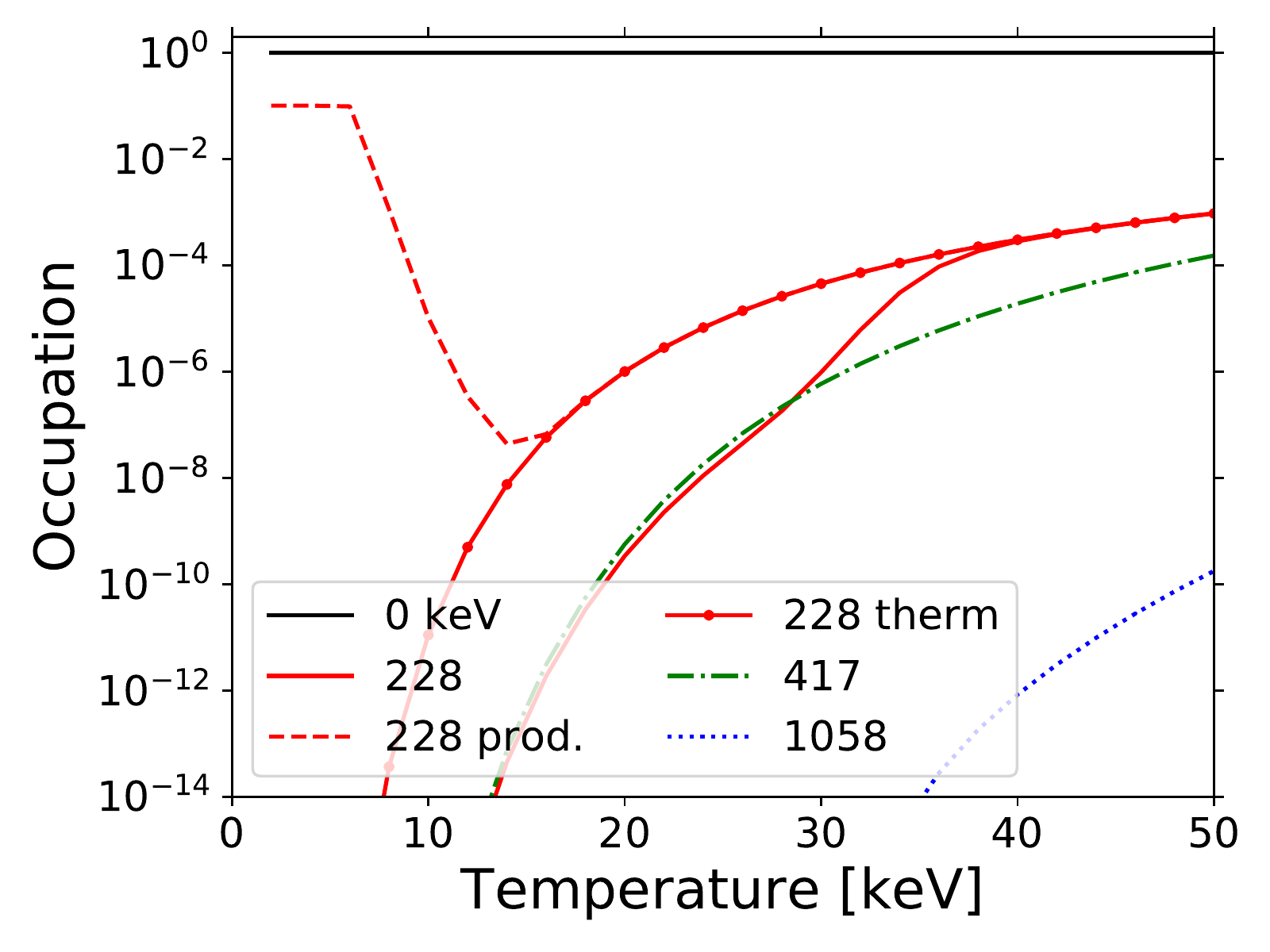}
    \caption{{\Al} occupation fractions $n_E$ of the lowest four levels.  The red lines show the isomer in steady state in the absence of production (solid), in steady state with production directly into the isomer (dashed), and in thermal equilibrium (circles).}
    \label{fig:al_occupation}
\end{figure}

Figure \ref{fig:al_occupation} shows the no-production ($\mathbf{P}=0$) steady state occupations $n_E$ of the lowest four levels as functions of temperature, as well as the steady state $n_{228}$ when {\Al} is produced in the isomeric state ($p_{228}=1$, $p_{E\neq 228}=0$).  The TE occupation of the isomer is included for comparison.

\begin{figure}
    \centering
    \includegraphics[width=\columnwidth]{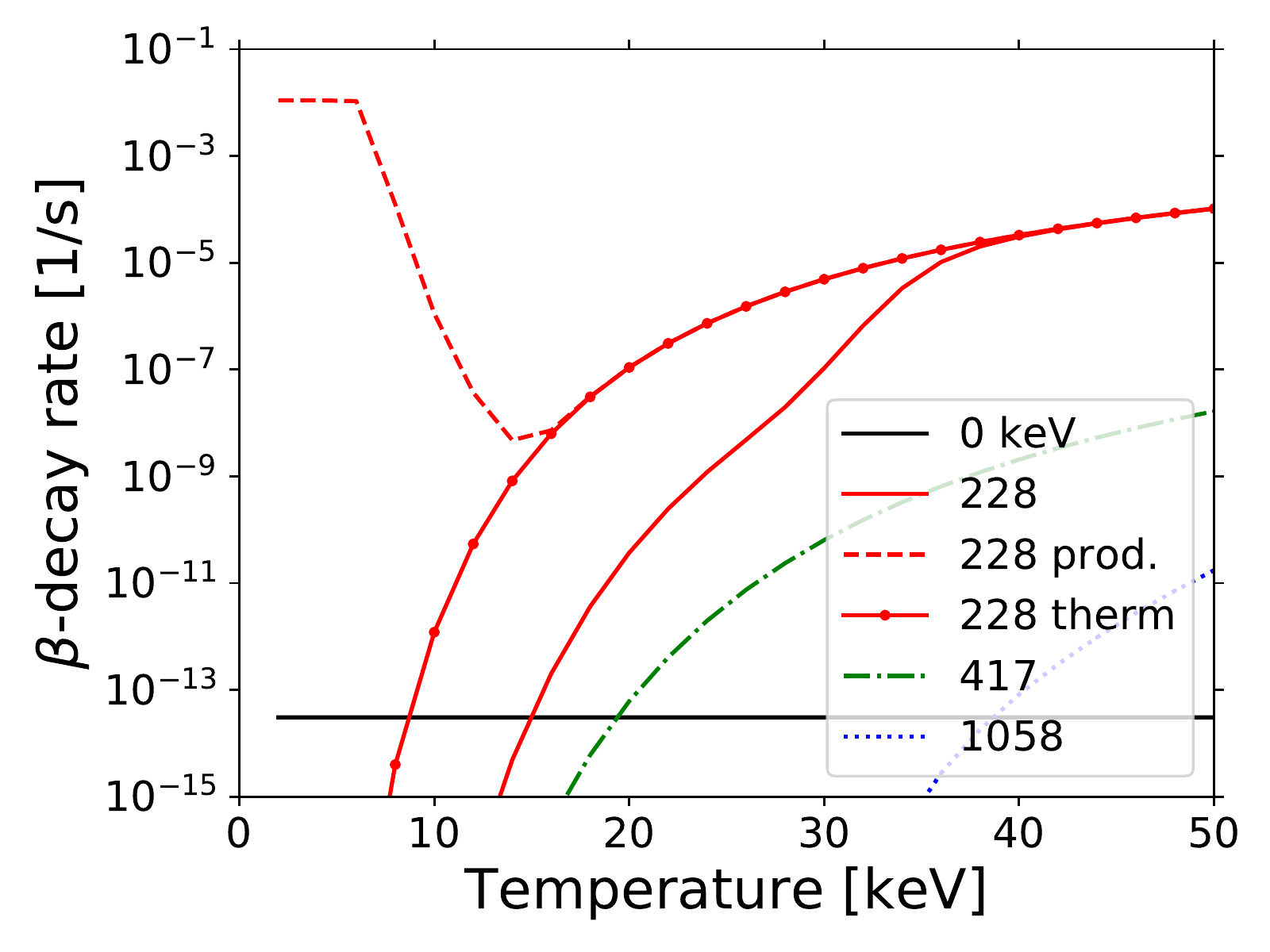}
    \caption{{\Al} individual contributions $n_E\lambda_E^\beta$ to the total $\beta$-decay rate $\lambda^\beta$ from the lowest four levels.  The red lines show the isomer in steady state in the absence of production (solid), in steady state with production directly into the isomer (dashed), and in thermal equilibrium (circles).}
    \label{fig:al_beta_ind}
\end{figure}

Folding the occupations $n_E$ with the $\beta$-decay rates $\lambda_E^\beta$ yields the contribution of each level to $\lambda^\beta$.  While the GS comprises the vast majority of the occupation at all temperatures shown here ($n_0\approx 1$), figure \ref{fig:al_beta_ind} makes clear that above $T=15$ keV, the isomer dominates the GS decay rate when $\mathbf{P}=0$.  If the isotope is produced in the isomeric state, then it dominates at all temperatures.  Comparison with the thermal value of $n_{228}\lambda_{228}^\beta$ shows that the isomeric nature of the state delays its $\mathbf{P}=0$ dominance for several keV in temperature.

\begin{figure}
    \centering
    \includegraphics[width=\columnwidth]{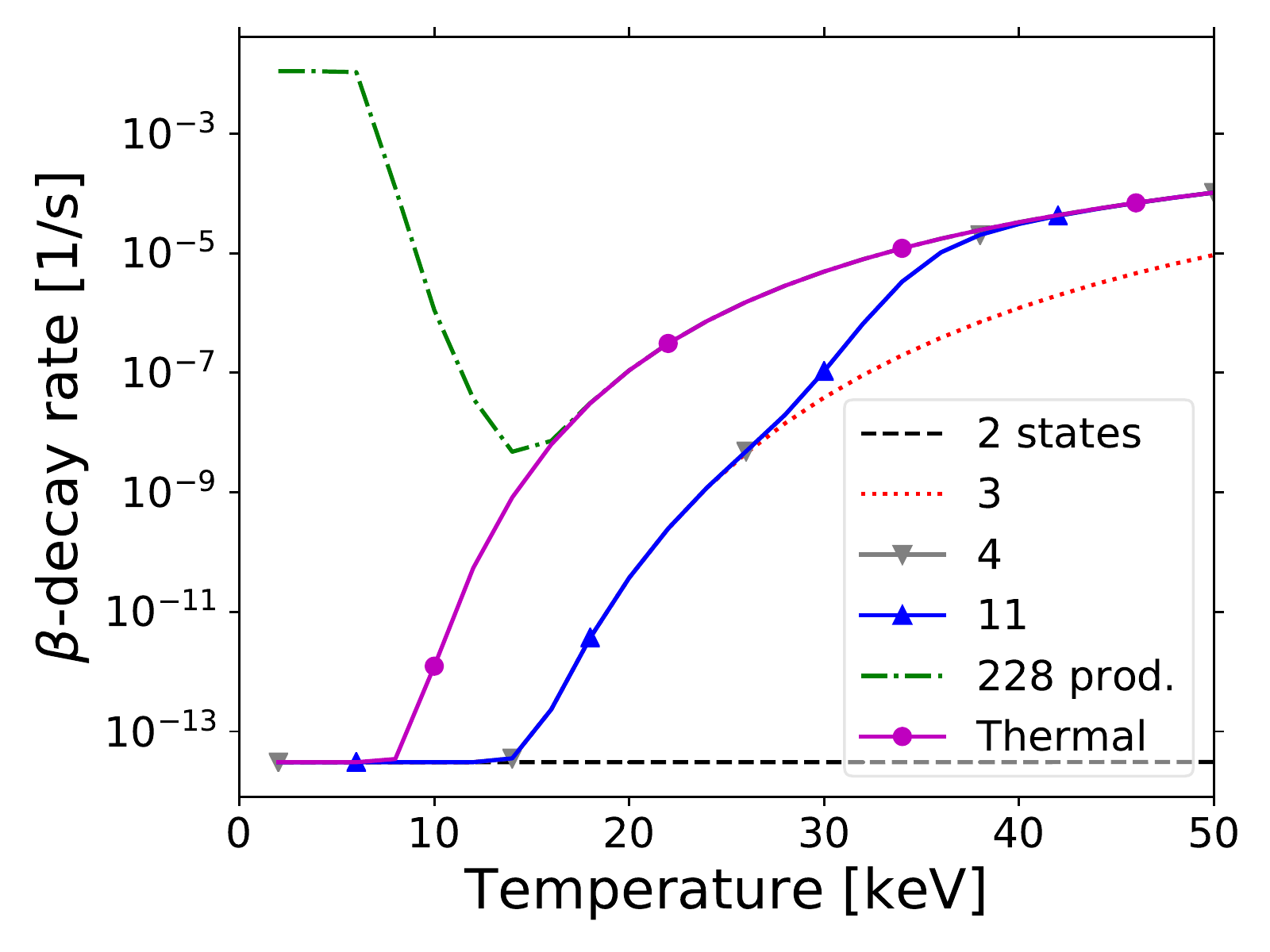}
    \caption{{\Al} total $\beta$-decay rates $\lambda^\beta$ under various assumptions.  The first four lines (black dashed, red dotted, gray downward triangles, blue upward triangles) show the steady state rates when 2, 3, 4, and 11 levels are included, respectively.  The fifth and sixth lines also include 11 levels; the fifth line (green dash-dotted) shows the rate with production directly into the isomer, and the sixth line (purple circles) shows the thermal equilibrium rate.}
    \label{fig:al_beta_total}
\end{figure}

Figure \ref{fig:al_beta_total} shows $\lambda^\beta$ as a function of temperature.  Steady state rates are shown for $\mathbf{P}=0$ when 2, 3, 4, and 11 nuclear levels are included in equation \ref{eq:ss_no_prod}.  Also shown are the rates for 11 included states with production $p_{228}=1$ and with the assumption of TE.  Because transitions directly between ground and the isomer are heavily suppressed, the 2 state calculation becomes wildly inaccurate where the isomer should be the major contributor to $\lambda^\beta$ ($T\gtrsim 15$ keV).  At $T\approx 30$ keV, the 3 state line diverges from the 11 state line, while the 4 state line stays in lock step with the 11 state line at all temperatures; this indicates that three levels are not sufficient to accurately describe {\Al}, but four are.

\begin{figure}
    \centering
    \includegraphics[width=\columnwidth]{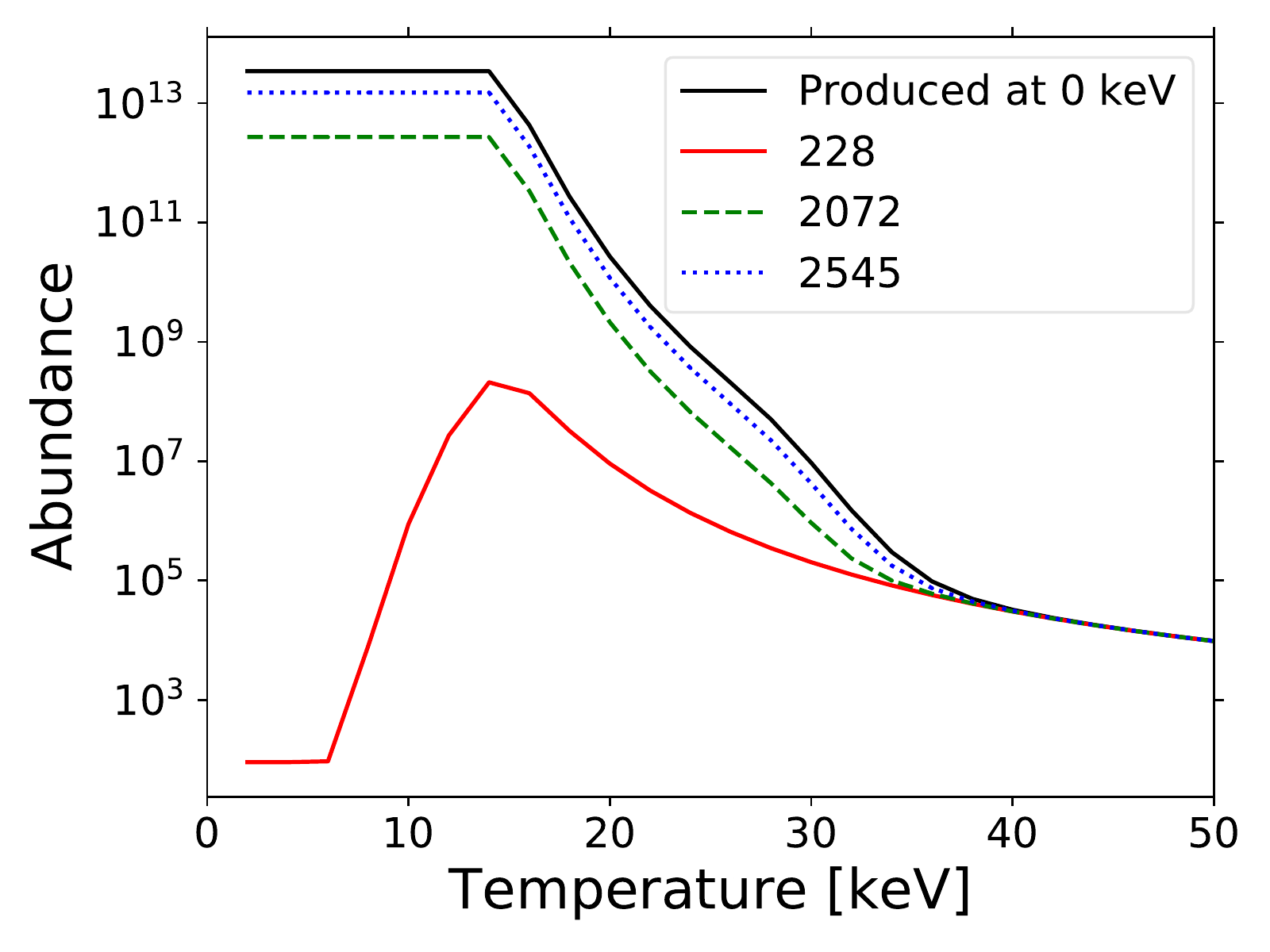}
    \caption{{\Al} steady state total abundance with constant production in the ground state (solid black), the isomer (solid red), at 2072 keV (dashed green), and at 2545 keV (dotted blue).  The production in all cases is normalized to 1 nucleus per second.}
    \label{fig:al_abundance_ss}
\end{figure}

Equation \ref{eq:P_equals_D} gives the steady state abundance $N$ of {\Al} when $\mathbf{P}\neq 0$.  Figure \ref{fig:al_abundance_ss} shows $N$ when $P_{E_i}=1$ s$^{-1}$, $P_{E\neq E_i}=0$ for $E_i=0$, $228$, $2072$, and $2545$ keV.  At moderate temperatures, even production directly into the isomer can eventually generate an appreciable abundance of the nuclide, as thermal transitions gradually build up a population in the GS.  Production into higher-lying states gives various results that depend principally on the branching ratio of $\gamma$ cascades to the ground and isomeric states.

\begin{figure}
    \centering
    \includegraphics[width=\columnwidth]{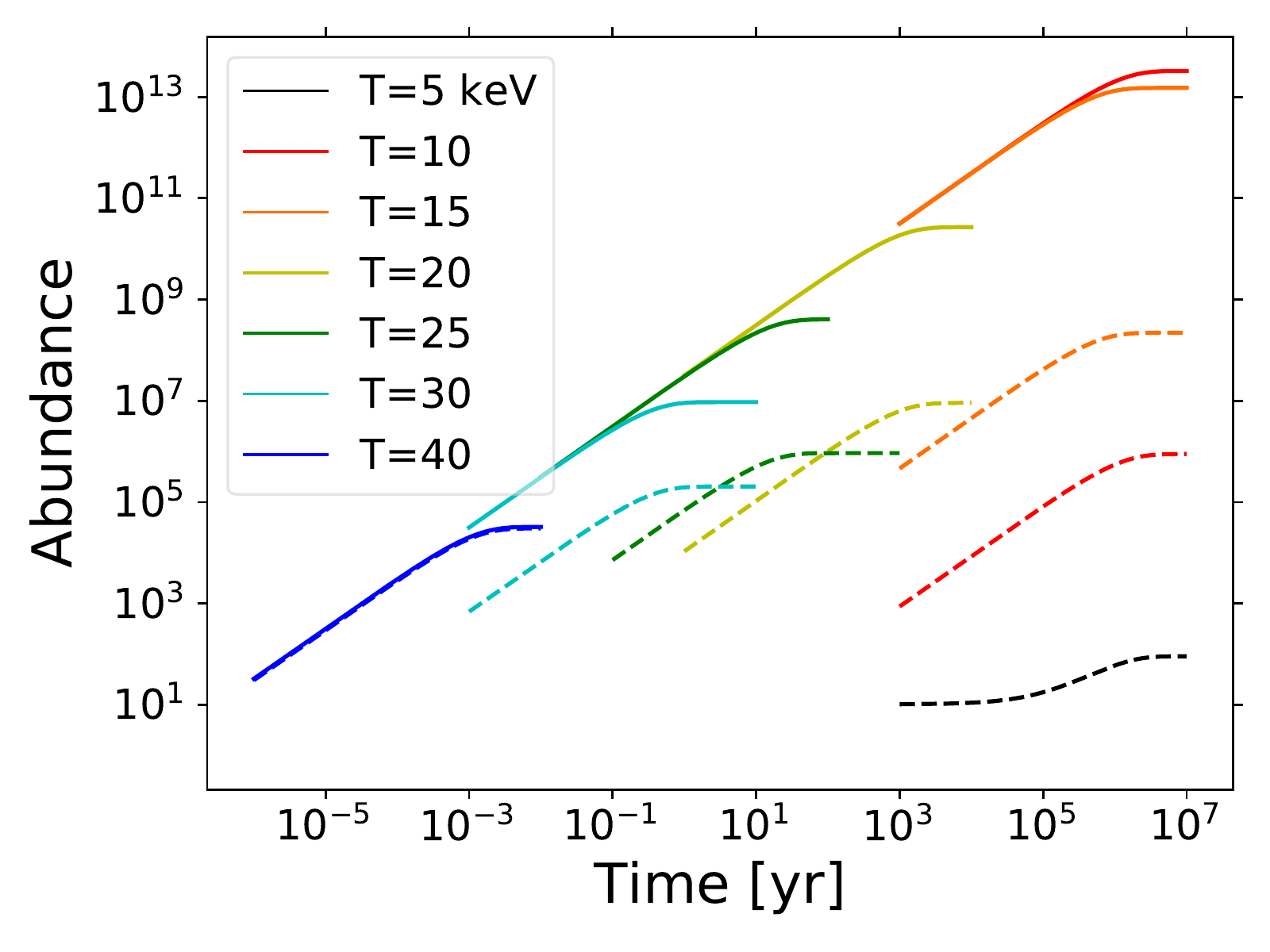}
    \caption{{\Al} total abundance as a function of time and temperature with constant production in the ground state (solid lines) and the isomer (dashed lines).  The production in all cases is normalized to 1 nucleus per second.}
    \label{fig:al_abund_dyn}
\end{figure}

Naturally, in most applications, the environment is time-dependent, so it is necessary to have a sense of the time required to reach the steady state abundance.  Figure \ref{fig:al_abund_dyn} shows $N$ as a function of time for production into the GS and production into the isomer at several temperatures.  Note that these are the steady state \emph{abundances}; in the absence of production, the occupation vector $\mathbf{n}$ rapidly converges from a thermal distribution to the steady state value.

All of these comparisons indicate that below $T\approx 40$ keV, thermal processes are slow enough that the $\beta$ decay of the isomer causes it to be depopulated relative to TE, with corresponding consequences for $\lambda^\beta$ and the abundance.  However, the calculations converge with TE at $T\approx 40$ keV; we conclude that above this equilibration temperature, TE rates are accurate.

\subsection{\Kr}

{\Kr} is a branch-point nucleus in the $s$-process.  Once created, it has two possible fates: $\beta$-decay to $^{85}$Rb, or capture a neutron to become $^{86}$Kr, which is $\beta$-stable.  Which path is preferred affects the final s-process abundance pattern, yielding information about the nucleosynthesis site.  {\Kr} $\beta$-decay properties are thus key to correctly interpreting observed abundance patterns.

{\Kr} has a long-lived isomer at 305 keV with $\lambda_{305}^\beta = 3.4 \times 10^{-5}$ s$^{-1}$.  While this is much faster than the the GS rate $\lambda_0^\beta = 2.0 \times 10^{-9}$ s$^{-1}$, the difference is not as extreme as in {\Al}.  Furthermore, due to both the longer $\beta$ lifetime of the isomer and the somewhat less forbidden IT directly to ground, the isomer has a $\sim 21\%$ branch to the GS as measured in the laboratory, i.e., at $T=0$.

Except where indicated, we carried out the calculations in this section using the lowest seven levels in {\Kr}.  We used experimental data whenever it was available.  We used the Weisskopf approximation for unmeasured spontaneous $\gamma$ transitions.  For unknown $\beta$ decays, we assumed $\log\left( ft\right)=5.0$ for every allowed daughter level.  The 1107 keV level is spin degenerate, and each of our calculations that uses that energy level includes both spins.  The spin of the 1167 keV level is experimentally uncertain; to compute the $\gamma$ and $\beta$ rates, we assigned the value $3/2$, which is within the experimental range.  As we will show, this and higher lying levels are not necessary to accurately describe the $\beta$-decay of {\Kr}, so we do not further address the spin uncertainty.  The input nuclear data is summarized in table \ref{tab:kr_nuc_data}.

\begin{table*}
\centering
\caption{Nuclear data used in our calculations of {\Kr}.  The columns are the initial state energy $E_i$, initial state spin and parity $J_i\pi$, $\beta$-decay rate $\lambda_i^\beta$, final state energy $E_f$, final state spin and parity $J_f\pi$, transition energy $E_\gamma$, multipole order of the electromagnetic transition, and spontaneous electromagnetic transition rate $\lambda_{if}^s$.  Starred (*) rates are experimental values.  All experimental data were taken from \cite{ENSDF}.}
\label{tab:kr_nuc_data}
\begin{tabular}{llr|lllcr}
\hline\hline
$E_{i}$ (keV)  & $J_{i}\pi$  & $\lambda_i^\beta$ (s$^{-1}$) & $E_{f}$ (keV)  & $J_{f}\pi$  & $E_\gamma$ (keV) & Multipolarity & $\lambda_{if}^{s}$ (s$^{-1}$) \\
\hline
0.0     & 9/2+ & *2.05$\times 10^{-9}$ & --       & --    & --       & --  & --                     \\
304.871 & 1/2- & *3.39$\times 10^{-5}$ & 0.0      & 9/2+  &  304.871 & $M4$ & *9.11$\times 10^{-6}$ \\
1107.32 & 1/2- & 1.70$\times 10^{-3}$  & 0.0      & 9/2+  & 1107.32  & $M4$ & 7.93$\times 10^{-2}$  \\
        &      & --                    & 304.871  & 1/2-  &  802.45  & $M1$ & 1.63$\times 10^{13}$  \\
1107.32 & 3/2- & 2.88$\times 10^{-3}$  & 0.0      & 9/2+  & 1107.32  & $E3$ & 6.70$\times 10^{5}$   \\
        &      & --                    & 304.871  & 1/2-  &  802.45  & $M1$ & 1.63$\times 10^{13}$  \\
        &      & --                    & 1107.32  & 1/2-  &       0  &   -- & 0                     \\
1140.73 & 5/2+ & 7.55$\times 10^{-5}$  & 0.0      & 9/2+  & 1140.73  & $E2$ & *1.98$\times 10^{11}$ \\
        &      & --                    & 304.871  & 1/2-  &  835.86  & $M2$ & 1.94$\times 10^{8}$   \\
        &      & --                    & 1107.32  & 1/2-  &   33.41  & $M2$ & 1.98$\times 10^{1}$   \\
        &      & --                    & 1107.32  & 3/2-  &   33.41  & $E1$ & 8.15$\times 10^{10}$  \\
1166.69 & 3/2- & 3.35$\times 10^{-3}$  & 0.0      & 9/2+  & 1166.69  & $E3$ & 9.66$\times 10^{5}$   \\
        &      & --                    & 304.871  & 1/2-  &  861.82  & $M1$ & 2.02$\times 10^{13}$  \\
        &      & --                    & 1107.32  & 1/2-  &   59.37  & $M1$ & 6.59$\times 10^{9}$   \\
        &      & --                    & 1107.32  & 3/2-  &   59.37  & $M1$ & 6.59$\times 10^{9}$   \\
        &      & --                    & 1140.73  & 5/2+  &   25.96  & $E1$ & 3.82$\times 10^{10}$  \\
1223.98 & 5/2- & 3.09$\times 10^{-3}$  & 0.0      & 9/2+  & 1223.98  & $M2$ & 1.31$\times 10^{9}$   \\
        &      & --                    & 304.871  & 1/2-  &  919.11  & $E2$ & *2.89$\times 10^{11}$ \\
        &      & --                    & 1107.32  & 1/2-  &  116.66  & $E2$ & 7.14$\times 10^{5}$   \\
        &      & --                    & 1107.32  & 3/2-  &  116.66  & $M1$ & 5.00$\times 10^{10}$  \\
        &      & --                    & 1140.73  & 5/2+  &   83.25  & $E1$ & 1.26$\times 10^{12}$  \\
        &      & --                    & 1166.69  & 3/2-  &   57.29  & $M1$ & 5.92$\times 10^{9}$   \\
        \hline\hline
\end{tabular}
\end{table*}

\begin{figure}
    \centering
    \includegraphics[width=\columnwidth]{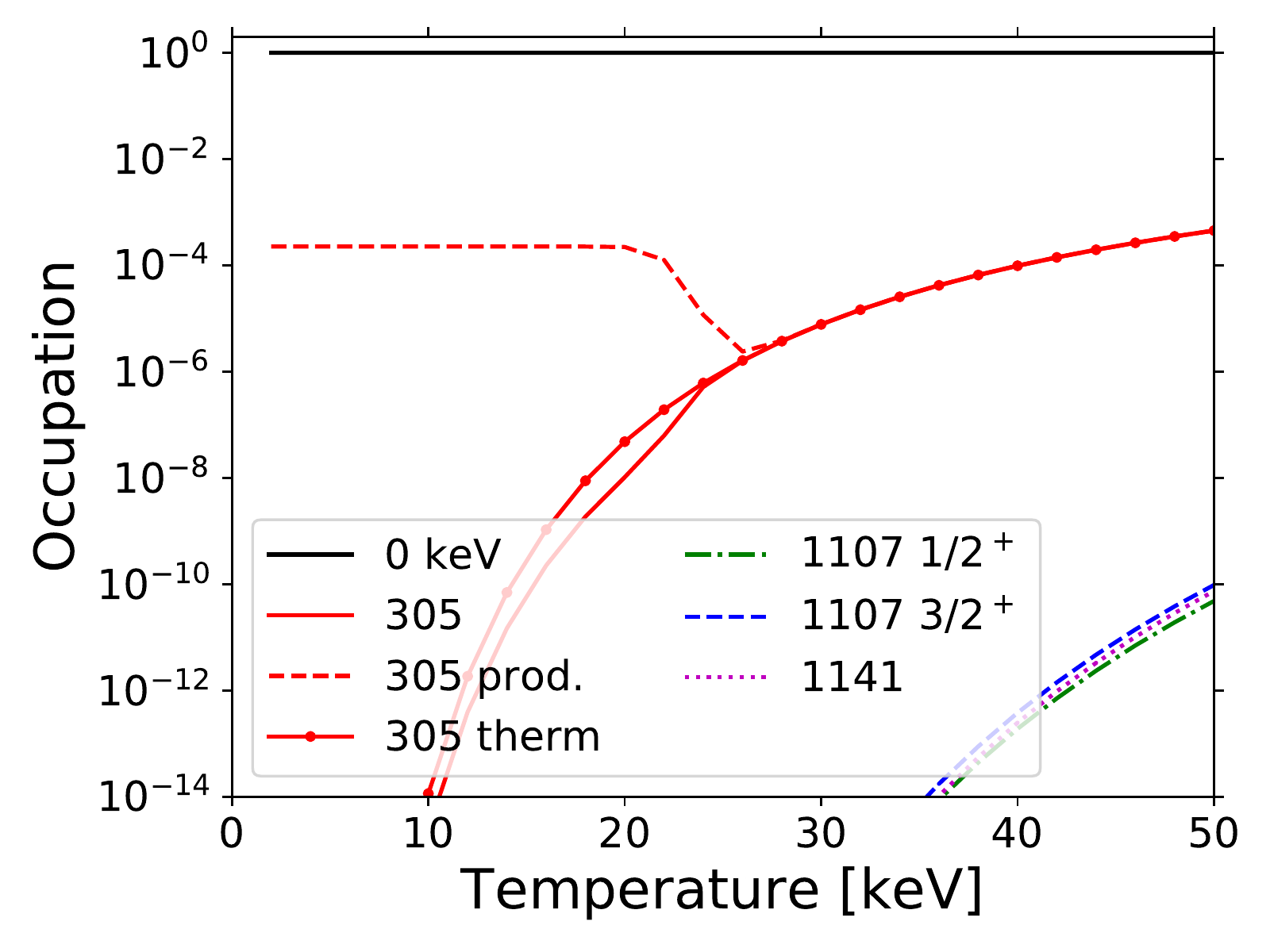}
    \caption{{\Kr} occupation fractions $n_E$ of the lowest five levels.  The red lines show the isomer in steady state in the absence of production (solid), in steady state with production directly into the isomer (dashed), and in thermal equilibrium (circles).}
    \label{fig:kr_occupation}
\end{figure}

Figure \ref{fig:kr_occupation} shows the steady state occupations $n_E$ for the lowest five levels in {\Kr} when $\mathbf{P}=0$, as well as the occupation of the isomer $n_{305}$ when $p_{305}=1$ and in TE.  As with {\Al}, the isomer becomes depopulated at low temperature, though the stronger coupling with the GS and the smaller difference in the ground and isomer $\beta$-decay rates reduce the effect.

\begin{figure}
    \centering
    \includegraphics[width=\columnwidth]{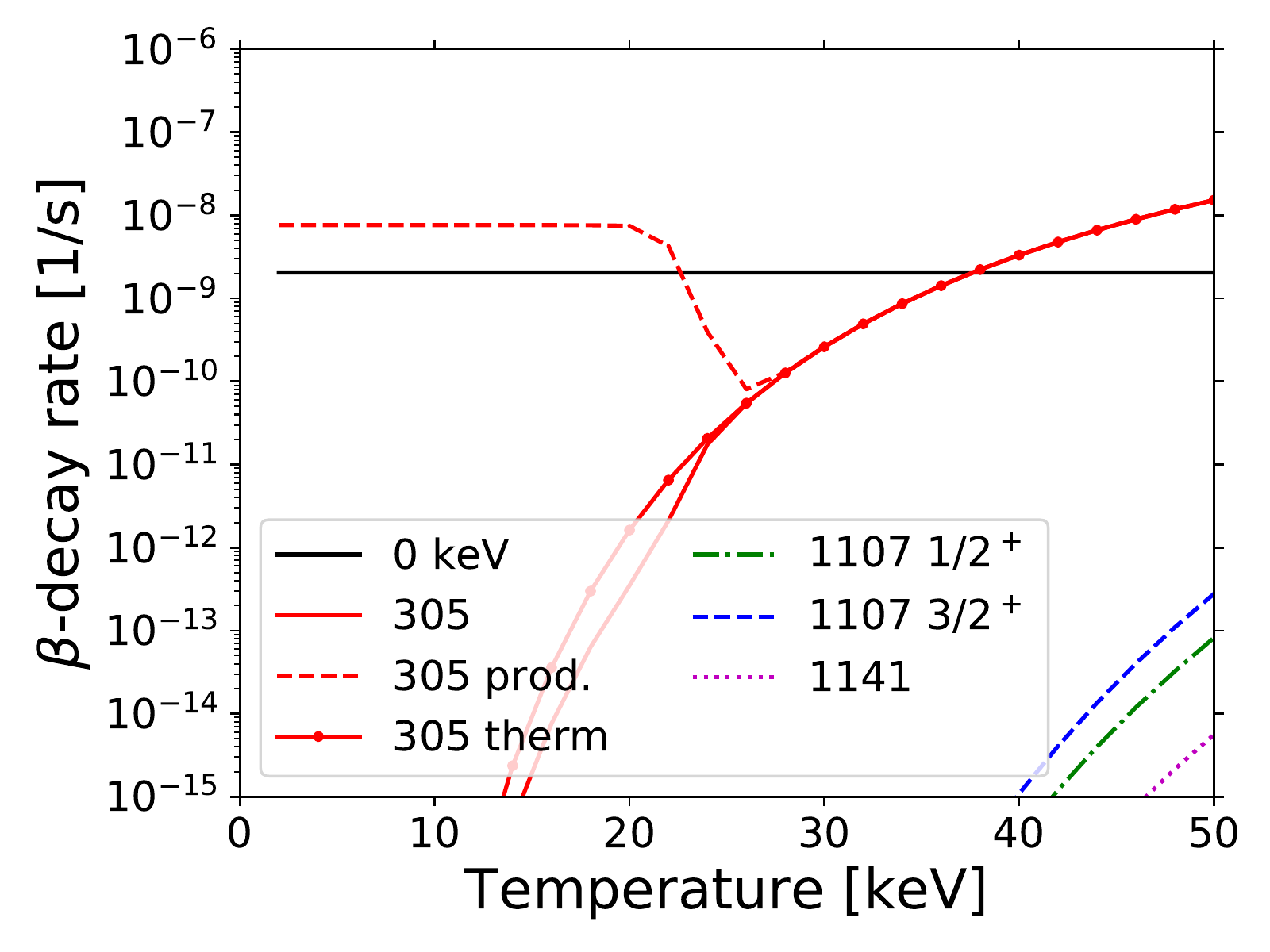}
    \caption{{\Kr} individual contributions $n_E\lambda_E^\beta$ to the total $\beta$-decay rate $\lambda^\beta$ from the lowest five levels.  The red lines show the isomer in steady state in the absence of production (solid), in steady state with production directly into the isomer (dashed), and in thermal equilibrium (circles).}
    \label{fig:kr_beta_ind}
\end{figure}

In contrast to {\Al}, the GS of {\Kr} dominates other contributions to $\lambda^\beta$ until well above the equilibration temperature of $\sim 25$ keV, as shown in figure \ref{fig:kr_beta_ind}.  On the other hand, production directly into the isomer ($p_{305}=1$) below the equilibration temperature still leads it to contribute dominantly to $\lambda^\beta$.

\begin{figure}
    \centering
    \includegraphics[width=\columnwidth]{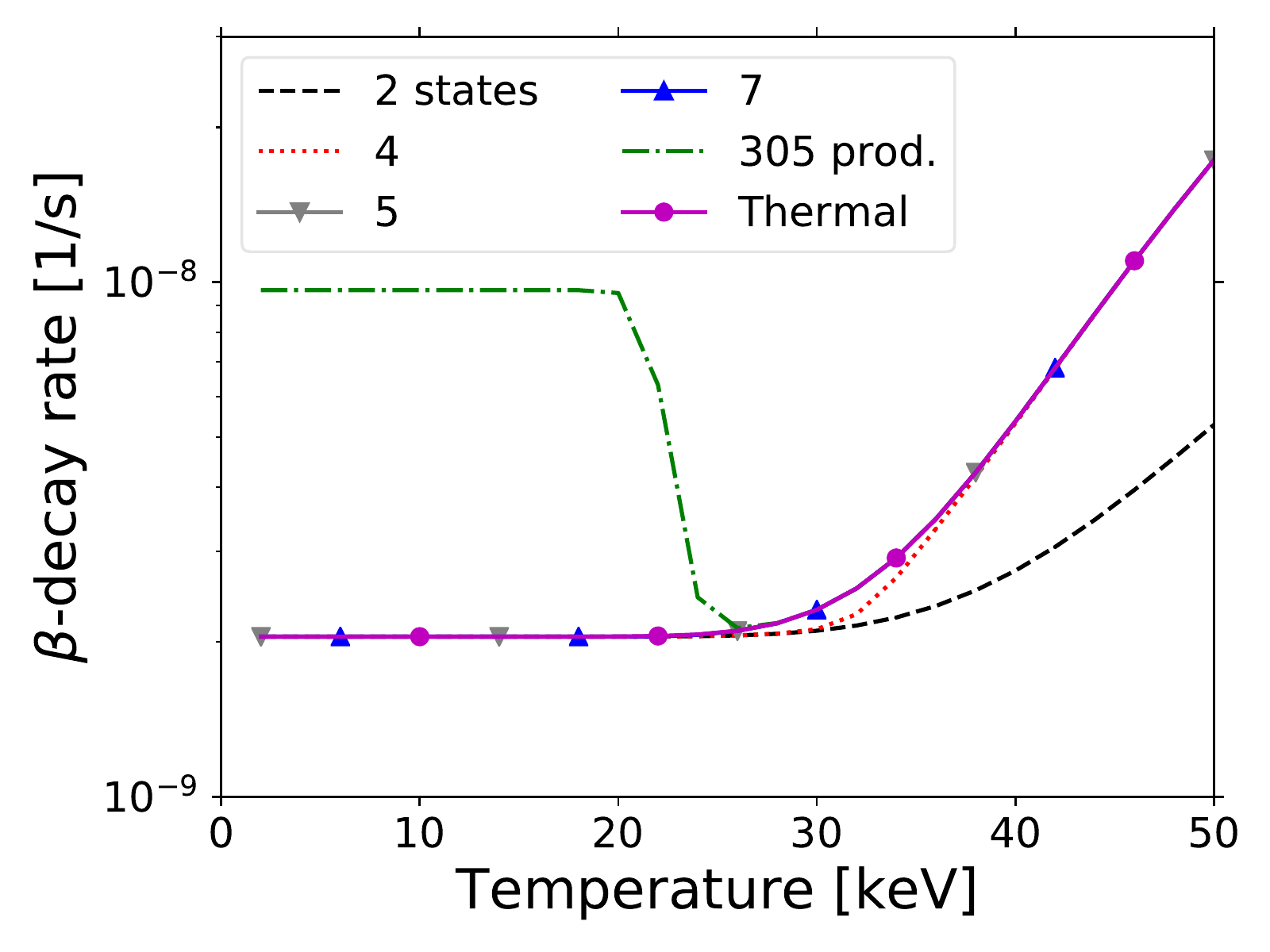}
    \caption{{\Kr} total $\beta$-decay rates under various assumptions.  The first four lines (black dashed, red dotted, gray downward triangles, blue upward triangles) show the steady state rates when 2, 4, 5, and 7 levels are included, respectively.  The fifth and sixth lines also include 7 levels; the fifth line (green dash-dotted) shows the rate with production directly into the isomer, and the sixth line (purple circles) shows the thermal equilibrium rate.}
    \label{fig:kr_beta_total}
\end{figure}

Figure \ref{fig:kr_beta_total} shows $\lambda^\beta$ as a function of temperature.  Steady state rates are shown for 2, 4, 5, and 7 included nuclear levels with zero production.  Also shown are the rates for 7 states with production directly into the isomeric level and the TE rate.  In the region from $T\sim 25-40$ keV, the 4 state line differs from the 5 and 7 state lines, the latter two of which remain together at all temperatures.  The 5 and 7 state $\mathbf{P}=0$ rates track the thermal rate exceedingly well, which in turn agrees with the calculations of \cite{ty:1987}.  Therefore, unless the {\Kr} production channel has a strong branch to the isomer, the TE rate is adequate.  If there is a major production branch to the isomer, five levels are necessary and sufficient to accurately describe {\Kr}.

\subsection{\Cl}

{\Cl} is a short-lived radioisotope.  Its GS $\beta$-decay rate is $4.5\times 10^{-1}$ s$^{-1}$, and it has an isomer at 146 keV with a $\beta$-decay rate of $2.0\times 10^{-4}$.  These rapid decays imply that {\Cl} can only be observed briefly after it is produced; nova outbursts provide such an opportunity \citep{lc:1987,endt:1990,coc2000}.

\begin{figure}
    \centering
    \includegraphics[width=\columnwidth]{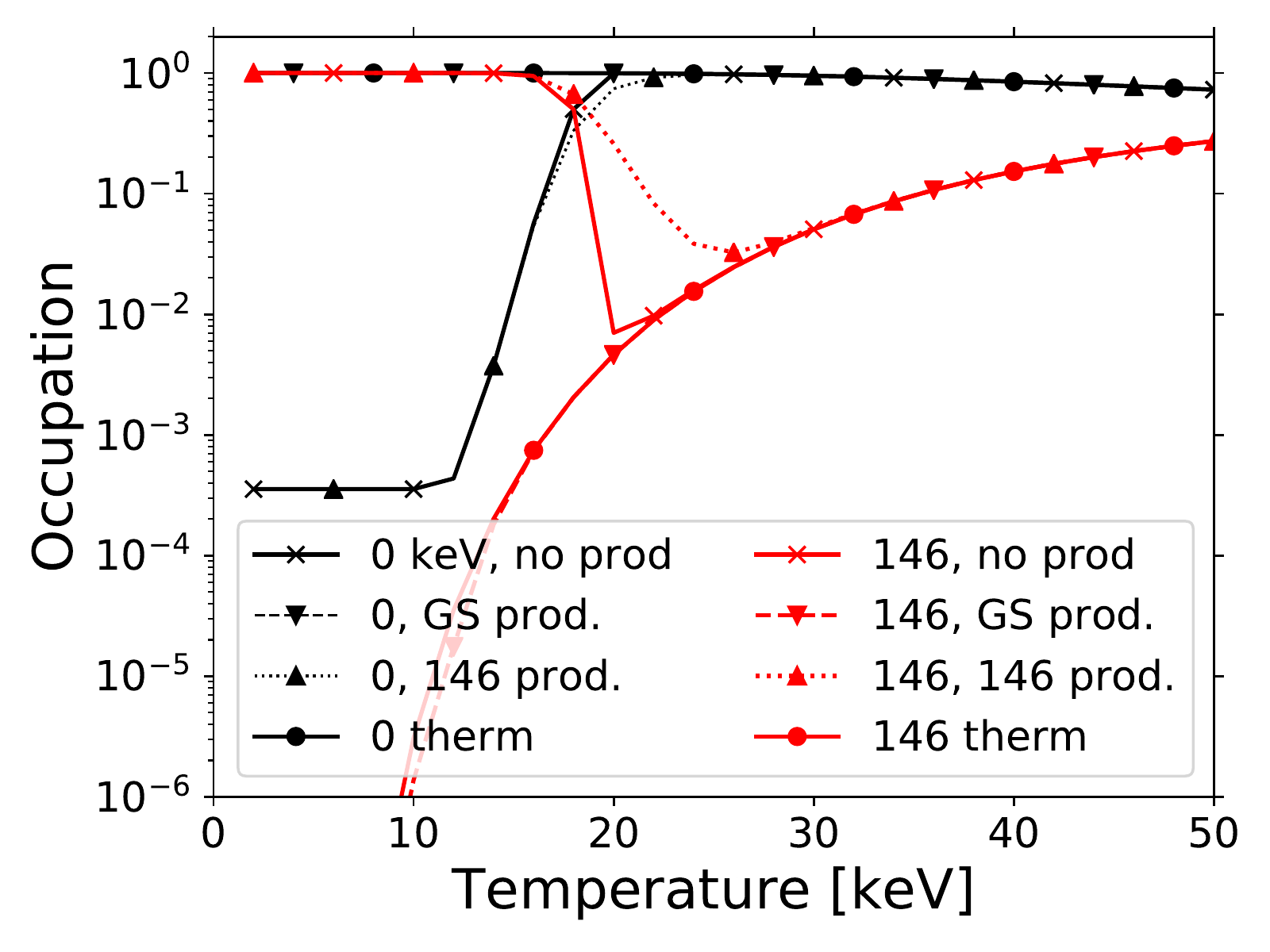}
    \caption{{\Cl} occupation fractions $n_E$ of the lowest two levels.  Black lines show the ground state, and red lines show the isomer.  Solid X lines show the results in the absence of production, dashed downward triangles show with production directly into the GS, dotted upward triangles show with production into the isomer, and solid circles show thermal equilibrium.}
    \label{fig:cl_occupation}
\end{figure}

Except as otherwise stated, we calculated {\Cl} using the lowest six nuclear levels.  Figure \ref{fig:cl_occupation} shows the steady state occupations of the ground and isomeric states in {\Cl}.  Because the isomer is more $\beta$-stable than the GS, the relationship between them in steady state is rather different from the cases of {\Al} and {\Kr}.  Whereas $n_0\approx 1$ in the previous two nuclei at all temperatures computed here, figure \ref{fig:cl_occupation} indicates that below the equilibration temperature of $20-25$ keV, the {\Cl} GS becomes depopulated.  Figure \ref{fig:cl_occupation} also shows $n_0$ and $n_{146}$ when $p_0=1$, when $p_{146}=1$, and in TE.  Below the equilibration temperature, production into either long-lived state causes it to comprise the bulk of the isotope's abundance.

\begin{figure}
    \centering
    \includegraphics[width=\columnwidth]{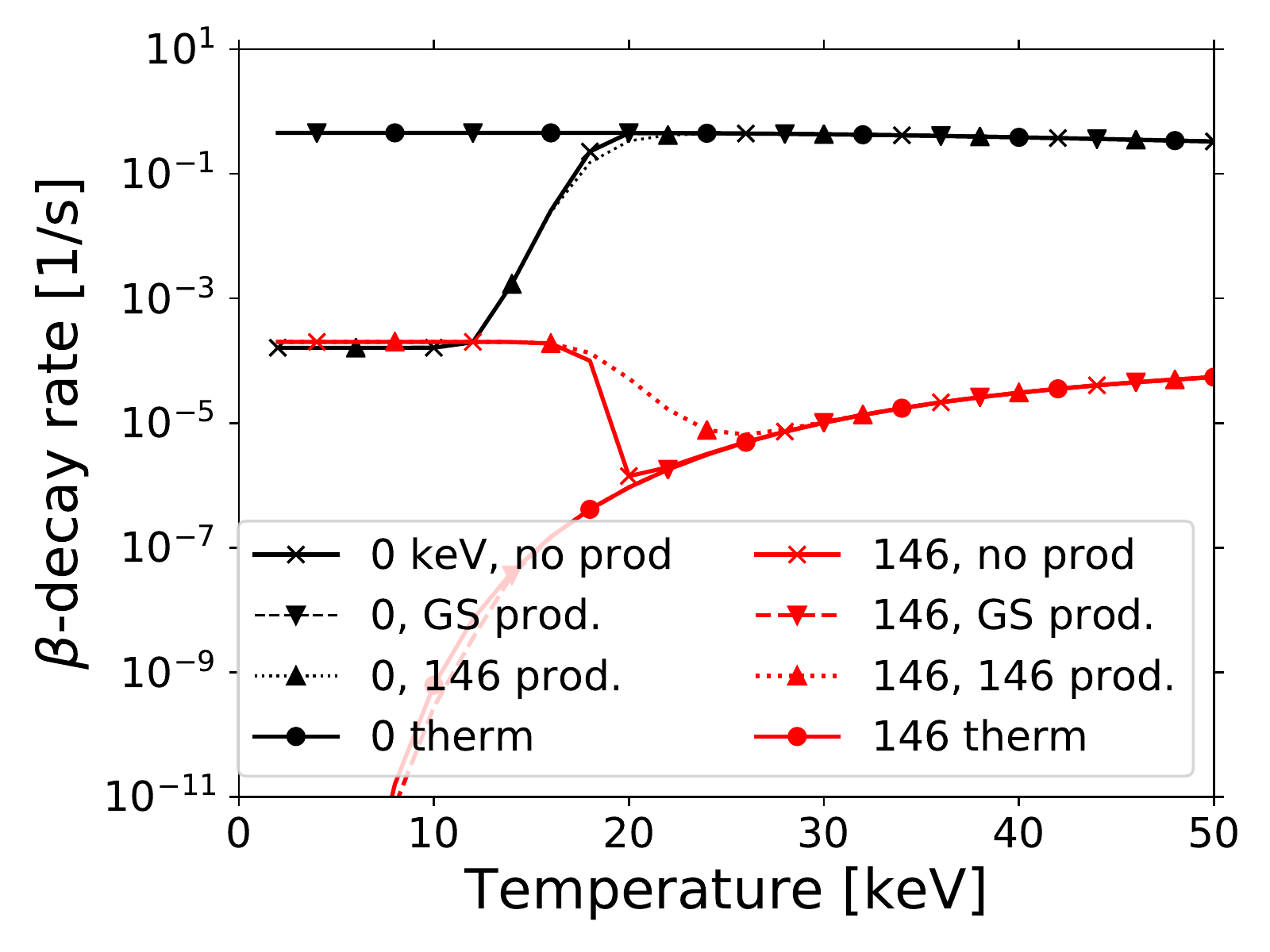}
    \caption{{\Cl} individual contributions $n_E\lambda_E^\beta$ to the total $\beta$-decay rate $\lambda^\beta$ from the lowest two levels.  Black lines show the ground state, and red lines show the isomer.  Solid X lines show the results in the absence of production, dashed downward triangles show with production directly into the GS, dotted upward triangles show with production into the isomer, and solid circles show thermal equilibrium.}
    \label{fig:cl_beta_ind}
\end{figure}

Folding the $\lambda_E^\beta$ with the $n_E$, shown in figure \ref{fig:cl_beta_ind}, reveals that regardless of the production channel, the GS contributes equally or dominates the isomer in $\lambda^\beta$.

\begin{figure}
    \centering
    \includegraphics[width=\columnwidth]{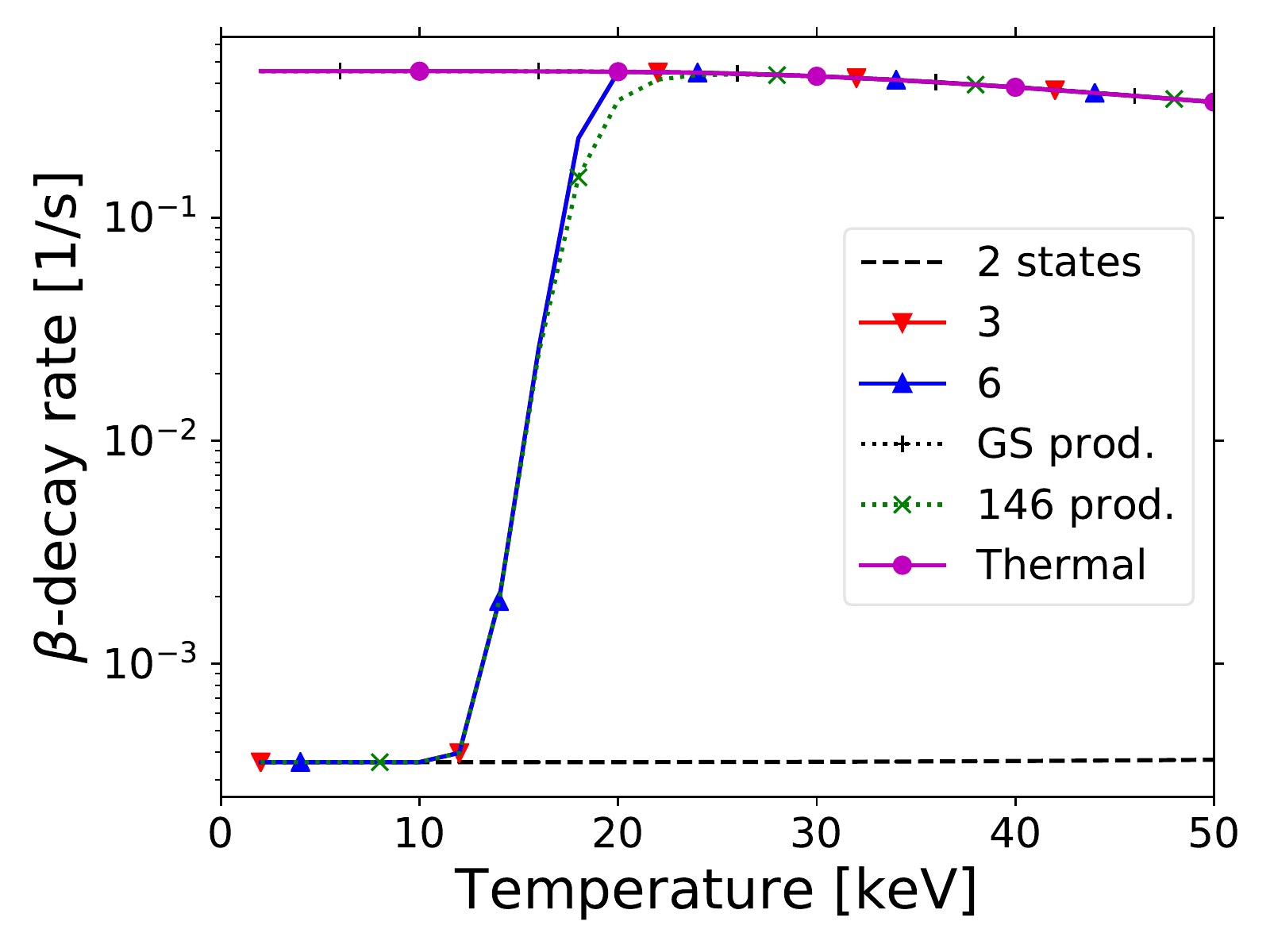}
    \caption{{\Cl} total $\beta$-decay rates under various assumptions.  The first three lines (black dashed, red downward triangles, blue upward triangles) show the steady state rates when 2, 3, and 6 levels are included, respectively.  The fourth, fifth, and sixth lines also include 6 levels; the fourth line (black dotted $+$) shows the rate with production directly into the ground state, the fifth line (green dotted X) shows the rate with production into the isomer, and the sixth line (purple circles) shows the thermal equilibrium rate.}
    \label{fig:cl_beta_total}
\end{figure}

Figure \ref{fig:cl_beta_total} shows $\lambda^\beta$ for {\Cl} with $\mathbf{P}=0$ when 2, 3, and 6 levels are included, as well as the rates when $p_0=1$, when $p_{146}=1$, and the thermal rate.  These results show that above $T\approx 25$ keV, the thermal rates apply regardless of production.  Below this equilibration temperature, three states are required for accurate results, and including more states does not improve precision.

\section{Discussion and Conclusions \label{sec:concl}}

We have described a method for estimating destruction rates of nuclei with long-lived isomers by computing their steady state behavior, and we have provided examples using $\beta$ decay.  While the time dependence of astrophysical environments will generally render the steady state approach inapplicable directly in nucleosynthesis network codes, it is nonetheless useful for evaluating other techniques, determining the minimum number of states other techniques must include (whether directly or indirectly), and the temperature above which thermal equilibrium rates apply (the equilibration temperature).

We emphasize that TE rates eventually become necessary.  Higher-lying states may contribute significantly to the total $\beta$-decay rate $\lambda^\beta$, and in such cases, any technique which treats $\beta$ decay as proceeding principally from the long-lived states must be abandoned at high temperature.  In {\Al}, for example, at $T\gtrsim 800$ keV, the thermal population of the 1058 keV level is similar to the isomer, owing to its greater spin degeneracy (see equation \ref{eq:ni_therm}).  The higher level has isospin $\tau =1$, and its corresponding superallowed $\beta$ decay dominates the isomer's contribution to $\lambda^\beta$.  Therefore, we recommend using TE rates above the equilibration temperature.

Below the equilibration temperature, our steady state results show that nucleosynthesis calculations should include four states in {\Al}, while three states are sufficient for {\Cl}.  As previous authors have found, the ground state of {\Al} electromagnetically couples most strongly to the 417 keV state, while the isomer couples to the 1058 keV state.  Transitions between the two higher levels then effect communication between the ground and isomeric states.  From table \ref{tab:kr_nuc_data}, the {\Kr} isomer couples to the $3/2^-$ 1107 keV level, with the 1141 keV level linking $3/2^-$ 1107 keV and ground.  While our calculations include all five lowest states, the inefficient transitions between $1/2^-$ 1107 keV and 1141 keV suggest that the former can be excluded.

In {\Cl}, the spontaneous $\gamma$-decay rate $\lambda_{42}$ of the 666 keV state (state 4) to the isomer (state 2) is approximately one order of magnitude faster than the experimental limit on the rate $\lambda_{32}$ from 461 keV (state 3) to the isomer; the calculations in this work used the shell model transition rate $\lambda_{32}$ reported in the appendix of \cite{banerjee-etal:2018}, which is more than two orders of magnitude slower than $\lambda_{42}$.  Furthermore, both the experimental limits and the shell model calculation of $\lambda_{32}$ indicate that it is a generically slow transition.  This qualitative analysis suggests that four states should be included.  However, the lower transition energy from the isomer to 461 keV renders it thermodynamically favorable which, coupled with the strong transition from 461 keV to ground, gives the result that three states are sufficient for {\Cl}.

Below the equilibration temperature, the total $\beta$-decay rates $\lambda^\beta$ are impacted significantly by which state a nuclide is produced in.  Therefore, we caution against using the bare no-production steady state $\lambda^\beta$ in nucleosynthesis network calculations.  While this is understood within the nucleosynthesis community, it bears repeating that network calculations should incorporate production channel branching into multiple long-lived states.

Our technique of solving for steady state occupations of energy levels in nuclei with isomers uses no theoretical approximation methods apart from estimates of experimentally unverified nuclear rates (internal transition and $\beta$-decay).  From the occupations, the overall rates of destruction processes such as $\beta$ decay can be computed for hot environments.  Comparing these occupations with thermal equilibrium calculations yields the equilibration temperature above which TE rates should be used.


We have shown (as previous authors have) that excited states facilitate transitions between the GS and isomer, and furthermore, more than one other excited state may be involved in the most efficient communication channels.  It is therefore prudent to ensure that the chosen technique employs sufficient states, whether they are included explicitly or implicitly; steady state calculations can confidently determine that number, since the occupations are sensitive to the efficiency of communication between the long-lived states.

Finally, we remark that the influence of long-lived isomers in astrophysics is likely insufficiently explored.  There are of course the issues with the nuclei discussed here, as well as with the cosmochronometer $^{182}$Hf \citep{lugaro-etal:2014} and others.  Furthermore, most nuclei in the nuclear chart are deformed, and K-isomers are common in well-deformed, heavy mass regions \citep{chen-etal:2012,kdk:2015,dwk:2016,wu-etal:2017}.  Thus, many nuclides in the r-process path have K-isomers, and there may be others in the s- and rp-process paths.  The structure issues of isomers and their consequences in astrophysics remain to be fully understood.

\section{Acknowledgments}

We thank Projjwal Banerjee for helpful discussions and insights.  This research was supported by the National Natural Science Foundation of China (No. 11575112) and the National Key Program for S\&T Research and Development (No. 2016YFA0400501).


\bibliography{references}

\end{document}